# Gas Sensing Properties of single-material SnP$_3$ logical junction via Negative Differential Resistance: Theoretical Study

Muhammad S. Ramzan,[†,‡] Han Seul Kim,[§, ¤,*] Agnieszka B. Kuc,[†,‡,*]

[†] Department of Physics and Earth Sciences, Jacobs University Bremen, Campus Ring 1, 28759 Bremen, Germany

[§] Center for Supercomputing Applications, National Institute of Supercomputing and Networking, Korea Institute of Science and Technology Information, 34141 Daejeon, Republic of Korea

[¤] Department of Data & High Performance Computing Science, University of Science & Technology, 34113 Daejeon, Republic of Korea

[‡] Helmholtz-Zentrum Dresden-Rossendorf, Abteilung Ressourcenökologie, Forschungsstelle Leipzig, Permoserstr. 15, 04318 Leipzig, Germany

*Supporting Information Placeholder*

**ABSTRACT:** The field of 2D materials has gained a lot of attention for vast range of applications. Amongst others, the sensing ability towards harmful gases is the application, which we explored in the present work using quantum-mechanical simulations for the SnP$_3$ material. Its electronic properties, namely 1 and 2 layers being semiconducting, while multilayers being metallic, offer a possibility to build a single-material logical junction. In addition, the harmful gases studied here show physical adsorption with charge transfer from the substrate to the gas molecules. Calculated recovery times show promise of a good sensing material. The I-V characteristics calculated for all cases indicates that SnP$_3$ could be a viable sensing material towards NO gas via negative differential resistance.

## INTRODUCTION

Industrialization and available means of transportation are continuously increasing the amount of toxic gases in our environment.[1] To control the indoor air quality,[2] flexible healthcare devices[3] and gas leak detectors,[4] sensors based on gas-sensing materials, are on high demand. Two-dimensional materials (2D) have recently been shown as very promising candidates for gas sensing applications, due to their novel electronic properties, high surface to volume ratio, and quantum effects, to name a few.[5–9] Graphene is considered as a state-of-the-art 2D material, however, its usage in electronic-based applications is very limited, due to the absence of an intrinsic bandgap.[10] There is, however, a wide variety of other 2D materials with semiconducting properties, available to date,[11–15] thus, they have been investigated, amongst others, in the direction of toxic gas detection.[4,5,16–18] Different 2D materials are sensitive to different gas species. For instance, field-effect transistors based on monolayer (1L) and multilayer MoS$_2$ were studied to detect NO gas. The multilayer MoS$_2$ devices were shown to be good candidates, however, for 1L device, the measured current upon exposure to NO gas was very unstable.[19] In another example, 1L C$_3$Si was shown as ideal sensing material for CO and NH$_3$ gases, while 1L boron-phosphorus seems to be an excellent sensor towards NO$_2$.[16,18] Additionally, theoretical studies showed that 1L MoSe$_2$ has higher sensitivity towards NO$_2$ among SO$_2$, NO, and NO$_2$ molecules,[20] whereas 1L WS$_2$ has an excellent sensing ability towards NO and O$_2$ gases.[21]

While the finite gap of 2D semiconductors makes them ideal for gas sensing, the bottleneck to their large-scale applications is the so-called Schottky barrier (SB), which builds up as semiconductors (channels) are brought in contact with metals (electrodes) to form a logical device.[22] In fact, a considerable amount of current dissipation at the channel-electrode interface causes, in general, a great hindrance in the realization of semiconductor-based electronic devices. Several approaches have been proposed to overcome this problem,[23–26] such as the ultralow contact resistance for conventional semiconductors by ion-implantation doping techniques, often abbreviated as I$^2$ techniques. However, I$^2$ techniques are not useful for low-dimensional semiconductors due to their atomic thickness. Recently, Andrews *et al.*[27] proposed an improved device contact with reduced SB by inserting additional semiconducting layers between 2D semiconductor channel (MoS$_2$) and metal (Ti/Au) electrodes.[27] However, this approach is relatively complex from the experimental standpoint.

Alternative approach would be to build a single-material device, where semiconducting channel and metallic electrodes constitute of the same material with different phases, overcoming the complexity and current drop at the electrode-channel interface. Kappera *et al.*[28] built a field-effect transistor (FET) with semiconducting 2$H$-MoS$_2$ as channel and locally induced metallic 1$T$–MoS$_2$ as electrodes.[28] The metastable nature of 1$T$ phase makes it, however, reconvert to 2$H$ phase at room temperature, hence, fundamentally limiting this approach. Ghorbani-Asl *et al.*[29] have proposed and simulated transport properties of a logical junction made of 2$H$-PdS$_2$ metallic multilayer electrodes and 1L semiconducting channel. This approach might be easier to achieve experimentally. Moreover, there has been emerging interest in such noble-metal materials for gas sensing devices,[29–31] for example, Teng-Yu *et al.*[31] have successfully synthesized PtSe$_2$-based gas sensor to detect NO$_2$ gas.[31] Cost of the noble-metal-based materials might be, however, a bottleneck for commercial use. More recently, others and we[12,13,32] have shown that SnP$_3$ or GeP$_3$ layered materials show similar quantum confinement effects as these in PdS$_2$ or PtSe$_2$, with 1L and 2L being semiconductors and multilayers being metals. Recent successful exfoliation of SnP$_3$ nanosheets using liquid-phase exfoliation techniques[33] has motivated us to further investigate SnP$_3$ as a potential single-material gas sensor.

In this work, we investigated from first-principles the potential of SnP$_3$ single-material device for gas sensing applications towards common pollutants: CH$_4$, CO$_2$, NH$_3$, NO, and NO$_2$. We found out that all studied molecules physically adsorb on the surface of 1L and 2L SnP$_3$, with weaker interaction for the latter. Based on the adsorption energy and corresponding recovery time, we divided studied molecules into two categories, type A (CH$_4$, CO$_2$, and NH$_3$) and type B (NO and NO$_2$). Transport properties were calculated using 3L SnP$_3$ as electrodes and 1L SnP$_3$ as transport channel. Transmission functions and *I*–*V* curves of such a device were analyzed towards sensing properties. Interestingly, in the range of applicable gate voltage, we observed the so-called negative differential resistance (NDR) for NO molecules, suggesting that SnP$_3$ is more sensitive towards this than other studied gases. Moreover, 2L SnP$_3$ shows a very good, calculated recovery time after exposure to NO gas. Our results potentially open up a new class of gas-sensing devices based on single materials, without presence of expensive components or junctions with large SB.

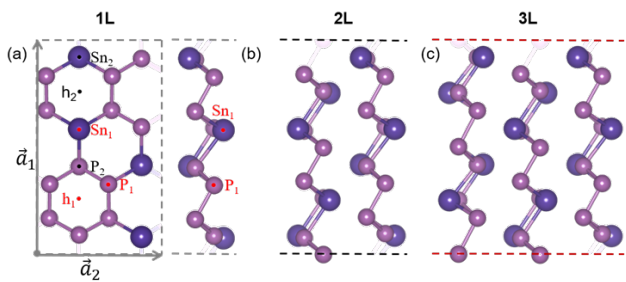

**Figure 1.** Atomic structures of 1L – 3L SnP$_3$ models in a rectangular unit cell representation with lattice vectors $\vec{a_1}$ and $\vec{a_2}$. Each individual layer is composed of two atomic layers (top and bottom) of both Sn and P atoms. (a) The possible adsorption sites on the top layer: h1 – middle of P hexagon, h2 – middle of P4Sn2 hexagon, P1 (Sn1) and P2 (Sn2) – on top of P (Sn) atom of the top (1) and bottom (2) atomic layers, respectively. In the 2L and 3L models, the layers are stacked in the AB and ABC manner, respectively.

**Computational Details.** Layered structures (1L, 2L, and 3L) were cut out from a bulk SnP$_3$, which has $R\bar{3}m$ space group.[34] In each layer of SnP$_3$, every Sn atom is attached to three P atoms, while each P atom is bound with one Sn and two other P atoms (see **Figure 1**). The layers are stack on top of one another in an ABC manner. Each individual layer forms a honeycomb pattern. All layered models were subsequently optimized (atomic positions and lattice vectors) as follows: 2L (***a*** = ***b*** = 7.255 Å) and 3L (***a*** = ***b*** = 7.283 Å) as hexagonal unit cells, while 1L (***a*** = 7.111, ***b*** = 12.318 Å) was modeled in the rectangular unit cell, to account for possible atomic reconstruction.[12] We found, however, no reconstruction in the latter. The adsorption of gas molecules and electronic transport calculations were performed on the rectangular supercells. As gas molecules, we considered CH$_4$, CO$_2$, NH$_3$, NO, and NO$_2$, and the calculations of adsorption sites, charge transfers, and charge density differences were performed on the 2 × 3 supercell. Various adsorption sites were considered (cf. **Figure 1a**): h$_1$ – middle of P hexagon, h$_2$ – middle of P$_4$Sn$_2$ hexagon, P$_1$ (Sn$_1$) and P$_2$ (Sn$_2$) – on top of P (Sn) atom of the top (1) and bottom (2) atomic layers, respectively. At each adsorption site, several representative molecular orientations were considered: horizontal and vertical or facing upward and downward for linear (CO$_2$ and NO) or non-linear (CH$_4$, NH$_3$, and NO$_2$) molecules, respectively (see **Figure S2**).

For geometries and electronic structure calculations, we used density functional theory as implemented in SIESTA (Spanish Initiative for Electronic Simulations with Thousands of Atoms) package.[35] We employed double-ζ plus polarization basis sets and Troullier-Martins type norm-conserving pseudopotentials[36] to describe core electrons, whereas exchange-correlation potential was treated within Perdew-Burke-Ernzerhof (PBE) functional of generalized-gradient-approximation (GGA).[37] We used $8 \times 16 \times 1$ ($6 \times 6 \times 1$) Monkhorst-Pack $k$-point mesh for unit cell (supercell) calculations. A vacuum of 30 Å along the direction perpendicular to the surface was used to avoid artificial interaction between periodic images.

The corresponding electronic band structures of 1L – 3L SnP$_3$ unit cells were calculated using optimized models (see **Figure S1**) and a good agreement with previous works was obtained: 1L and 2L are semiconductors, while 3L is metallic.[12,32]

The geometries were relaxed to $10^{-4}$ eV and 0.01 eV per Å for the electronic energy and total atomic force convergence, respectively. The adsorption energies were calculated as follows:

$$E_{ads} = E_{SnP_3+gas} - E_{SnP_3} - E_{gas} - E_{cc} \quad (1)$$

where $E_{ads}$ denotes the adsorption energy, while $E_{SnP_3+gas}$, $E_{SnP_3}$, and $E_{gas}$ are the total energies of the system, the substrate, and the gas molecule, respectively. $E_{cc}$ refers to the counterpoise correction, which may occur due to incompleteness of the basis sets.[38]

The single-material device was constructed from metallic left and right electrodes based on tri-layer SnP$_3$ (3L) and a scattering region based on 1L SnP$_3$. The difference between the lattice constants of 3L and 1L was about 2.4%, so we used that of the latter for the whole device, with no significant change in the electronic properties of the electrodes, which stay metallic. We used $2 \times 1$ supercell of 3L SnP$_3$ as electrodes, with additional $2 \times 1$ overlap region as a buffer, and $2 \times 5$ supercell of 1L SnP$_3$ as transport channel (hereafter, it will be referred to as a 313-model). For comparison, we also constructed a model with electrodes ($2 \times 3$ supercell) and transport channel ($2 \times 6$ supercell) made of 1L SnP$_3$ (111-model). Quantum transport calculations were performed using non-equilibrium Green's function (NEGF) theory as implemented in TranSIESTA code.[39] The transmission function, $T(E)$, is written as:

$$T(E,V) = Tr[\Gamma_L(E,V)G(E,V)\Gamma_R(E,V)G^\dagger(E,V)] \quad (2)$$

where $\Gamma_L$ ($\Gamma_R$), $G$, and $V$ are the broadening matrix of left (right) electrode, Green's function, and the bias voltage, respectively. The current-voltage ($I$-$V$) characteristics were obtained using TBtrans code based on the Landauer-Büttiker formula:[40]

$$I(V) = \frac{2e^2}{h} \int (f_L(E) - f_R(E)) T(E,V) dE, \quad (3)$$

where $e$, $h$, and $f$ are the electron charge, Planck's constant, and Fermi-Dirac distribution function, respectively. The applied bias, $V$, can be thought of as the chemical potential differences between the two electrodes divided by the electron charge, $\frac{\mu_L - \mu_R}{e}$. The Fermi-Dirac distribution function of electrode L/R is written as: $f_{L/R} = \frac{1}{e^{(E-\mu_{L/R})/k_B T}+1}$, where $k_B$ is the Boltzmann's constant and $T$ is the electronic temperature.

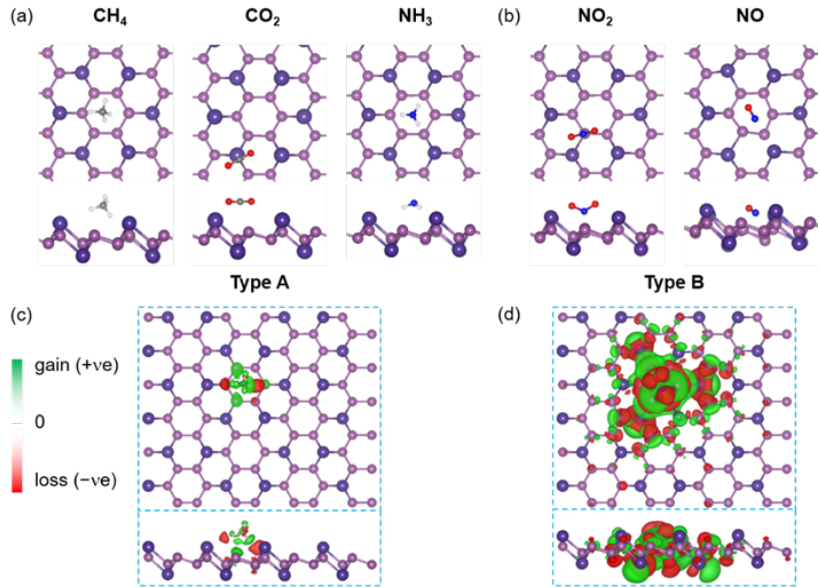

**Figure 2.** Top and side views of energetically most stable configurations of gas molecules of type A (a) and B (b) adsorbed on 1L SnP$_3$. Exemplary charge density differences (c-d) of type A (CH$_4$) and type B (NO) cases. Green and red areas represent charge accumulation and charge depletion, respectively, with isovalue of $3 \times 10^{-4}$ e/Å$^3$. Atoms color code: H – white, C – grey, O – red, N – blue, Sn – purple, and P – pink.

**Table 1.** Adsorption sites (see **Figures 1** and **S2** for details) and adsorption energies ($E_{ads}$) of gas molecules on 1L and 2L SnP$_3$. The charge transfer ($\Delta q^{1L}$) from 1L SnP$_3$ to an adsorbed molecule based on Mulliken charge analysis. The recovery times ($\tau$) for 1L and 2L cases.

| Gas | Site | $E_{ads}$ (eV) | | $\Delta q^{1L}$ (e$^-$) | $\tau$ | |
|---|---|---|---|---|---|---|
| | | 1L | 2L | | 1L | 2L |
| CH$_4$ | h$_1$ | -0.45 | -0.29 | 0.04 | 0.32 $\mu s$ | 8.68 $ns$ |
| CO$_2$ | Sn$_2$ (H) | -0.36 | -0.37 | 0.24 | 1.26 $\mu s$ | 138 $ns$ |
| NH$_3$ | h$_1$ (D) | -0.52 | -0.35 | 0.03 | 0.06 $\mu s$ | 73.5 $ns$ |
| NO$_2$ | P$_1$ (U) | -1.52 | -0.97 | 0.67 | 0.33 $Ts$ | 1.66 $Ms$ |
| NO | h$_1$ (V) | -1.33 | -0.76 | 0.59 | 0.19 $Ps$ | 70 $\mu s$ |

**Results and Discussion.** **Figures 2a-c** show the optimized geometries of gas molecules adsorbed on 1L SnP$_3$ and their corresponding adsorption energies ($E_{ads}$) are given in **Table 1**. We can divide all molecules into two types according to the $E_{ads}$: types A (CH$_4$, CO$_2$, and NH$_3$) – weak adsorption and B (NO and NO$_2$) – strong adsorption. In type A, the CO$_2$ molecule prefers horizontal orientation at Sn$_2$ site, while CH$_4$ and NH$_3$ adsorb on h$_1$ sites, with the latter molecule pointing its H atoms towards the SnP$_3$ layer (**Figure 2a**). In type B, NO$_2$ prefers the P$_1$ site with O atoms pointing upward, while NO favors h$_1$ site with nearly vertical orientation (see **Figure 2b**). The latter case also causes a small reconstruction in the SnP$_3$ layer, namely in the shape of the P-hexagon close to the adsorption site. Such a reconstruction was not observed for the other molecules. Based on the adsorption analysis, we could conclude that SnP$_3$ might be a suitable substrate for detecting N-based gases, especially NO molecules.

We have also investigated the adsorption of the same gas molecules on 2L SnP$_3$ and, consistent with previous reports,[41] we found that the additional layer weakens the interaction of molecules with the substrate in all cases except for CO$_2$. The additional layer, however, does not change the adsorption sites.

Next, we calculated the charge transfer ($\Delta q^{1L}$) between the 1L SnP$_3$ and the adsorbed molecules by means of Mulliken charge analysis (see **Table 1**) and the charge density difference (CDD; see **Figure 2c-d** and **Figure S3**). CDD is defined as:

$$CDD = \rho_{SnP_3+gas} - \rho_{SnP_3} - \rho_{gas}, \quad (4)$$

where $\rho_{SnP_3+gas}$, $\rho_{SnP_3}$, and $\rho_{gas}$ indicate the charge densities of 1L SnP$_3$ with adsorbed gas molecule, 1L SnP$_3$, and gas molecule, respectively. Mulliken analysis shows that the charge was transferred from the substrate to the gas molecule in each case. The CDD shows that for type A (type B) the overall charge densities are localized (delocalized) around the gas molecule. Based on the $\Delta q^{1L}$ values (**Table 1**), the molecules could again be divided into two groups: type A (CH$_4$, NH$_3$, and CO$_2$) – small or intermediate (for CO$_2$) charge transfer and B (NO and NO$_2$) – large charge transfer. Although the total amount of charge transferred between SnP$_3$ and CO$_2$ is larger than for other type A molecules, the CDD shows that the local change of charge distribution at the interface is more noticeable in the latter. In case of type B, the CCD shows strong delocalization of the transferred charge.

Another important property of a gas sensing material is its recovery time ($\tau$), which indicates the re-useability of the SnP$_3$ substrate. The recovery time can be calculated as follows:

$$\tau = \nu^{-1} exp \frac{|E_{ads}|}{k_B T}, \quad (5)$$

where $\nu$, k$_B$, and T are the attempt frequency of bond breaking, the Boltzman constant, and temperature, respectively, and $E_{ads}$ is the adsorption energy. One of the reference works showing a good recovery time of $\tau = 16\ s$ for $\nu = 1$ THz at room temperature is based on the simulations of $NO_2$ adsorption on carbon nanotubes.[42] For 2D materials, some experimental works showed $\tau = 150\ s$ for $NH_3$ on $WS_2$ flakes and $\tau = 240\ s$ for NO on $In_2O_3$ nanosheets.[43,44] We also adopted the same value of attempt frequency, $\nu = 1$ THz, in order to calculate the room temperature (T = 300 K) recovery times. Interestingly, our results show layer-dependent recovery time: for 1L $SnP_3$, type A molecules show ultrafast recovery of 0.32 $\mu s$, 1.33 $\mu s$, and 0.06 $\mu s$ for $CH_4$, $CO_2$ and $NH_3$, respectively. These values decrease for the case of 2L $SnP_3$, consistent with the change in the adsorption energies, and type A molecules enter the ns recovery time regime. Type B molecules show very long recovery times for 1L, which also strongly decreases for the 2L case. Especially, the recovery time for NO molecule becomes very good with $\tau = 70\ \mu s$ (**Table 1**). To desorb $NO_2$ at room temperature, the substrate would require heating. It is important to mention that 2L $SnP_3$ system with $\tau$ in the regime of ns towards type A molecule might not be a good sensor, as the recovery time is way too fast to detect anything. Thus, we could conclude from this analysis that 1L $SnP_3$ might be a good sensor for type A gas molecules, while 2L $SnP_3$ for NO.

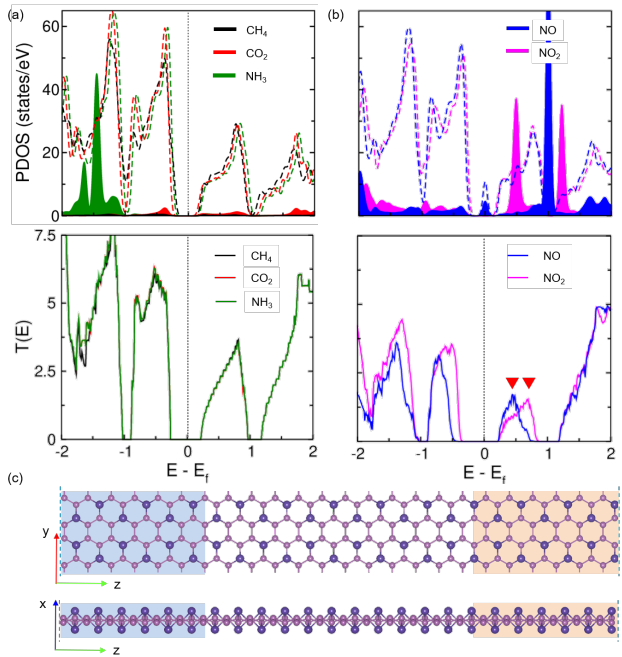

**Figure 3.** Projected density of states (PDOS) and the corresponding transmission (T(E)) of 1L $SnP_3$ with adsorbed gas molecules of type A (a) and type B (b). The states corresponding to each molecule in the PDOS are marked with color filled peaks and were scaled x10 for visibility. Dashed lines correspond to the substrate stated for each case. Note that the states of $CH_4$ lie outside the energy range plotted here. (c) The corresponding schematic of the device (111-model) used to calculate the transmission with blue and orange regions marking the leads.

Next, we analyzed electronic and transport properties of 1L $SnP_3$ substrate with adsorbed gas molecules. **Figure 3** show the projected density of states (PDOS) and the corresponding transmission function for all studied cases. The state contributions from the adsorbed molecules appear far from the band edges for $CH_4$ and $NH_3$, close to the band edges for $CO_2$ and $NO_2$, and at the Fermi level ($E_F$) for NO. The latter induces a metallic character of $SnP_3$ upon NO adsorption. For type B molecules, there is a hybridization between the molecular and substrate states (cf. **Figure 3b**).

The transmission (T(E)) of 1L $SnP_3$ upon adsorption of molecules was calculated at zero bias using the 111-model shown in **Figure 3c**. The values of T(E) indicate that molecules of type A induce only a negligible change to the intrinsic transport properties of the pristine 1L $SnP_3$ (cf. **Figure S4**), being consistent with the PDOS. Note that such negligible changes in T(E) are due to the insignificant hybridization between molecular and $SnP_3$ state. On the other hand, a noticeable reduction of T(E) over a wide energy range is observed for type B molecules, resulting from the strong hybridization of state. The effect is stronger for the NO molecule, not only showing lower T(E) values, but also the first peak above the Fermi level (indicated by red triangles) comes closer to $E_F$.

Finally, we investigated the gas sensing characteristic of a single-material logical junction. For this, we used our 313-model as shown in **Figure 4a** (see method section for details). We calculated the zero-bias T(E) and the I-V (current-voltage) characteristics for different bias values. In agreement with the cases of the 111-model device, small changes in the T(E) are observed for type A molecules (see **Figure 4b**). For type B, the strongest change is again visible for NO molecule, not only the T(E) is decreased, but there is a prominent shift of the first peak to lower energies compared to $NO_2$.

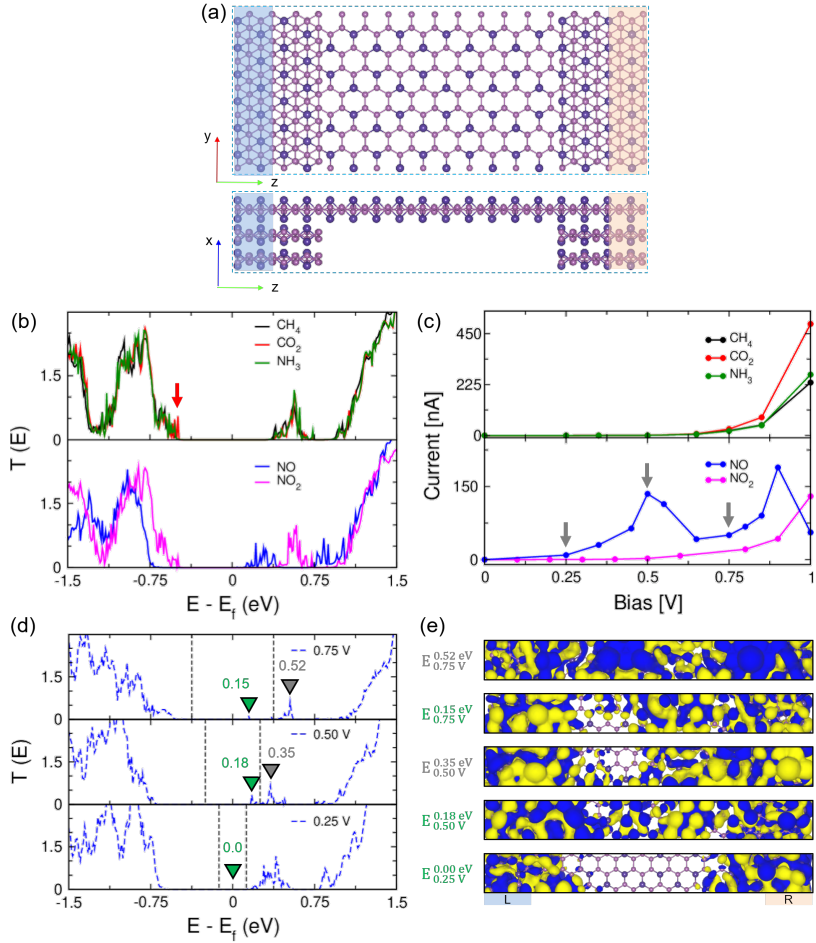

**Figure 4.** Transport calculations of a device with 1L SnP$_3$ transport channel and 3L SnP$_3$ leads: (a) the atomic structure of the device, with electrodes (highlighted with blue and orange rectangles), overlap region as a buffer, and the transport channel. (b) The transmission function for gas molecules adsorbed on top of the transport channel and (c) the corresponding I-V curves for positive bias voltage. The details of transmission calculations focused on the NO molecule, showing the negative differential resistance (NDR). (d) The transmission functions for three different bias values that are marked with the downward arrows on the I-V curve for NO molecule. For each bias, the filled green and grey triangles correspond to the transmission peaks inside and outside of the bias window, respectively. The eigenchannels for each triangle are plotted in (e).

In the most interesting case of NO molecule, the current increases first to a maximum value of about 150 nA at 0.5 V. It then decreases with increasing voltage to about 42 nA for 0.65 V, resulting in NDR. From that point on, the current again increases to another maximum at 0.90 V and then decreases again. To better understand the NDR for NO molecule, we analyzed the T(E) (see **Figure 4d**) and the corresponding scattering eigenchannels (see **Figure 4e**) at three representative voltage points, marked with downward arrows in **Figure 4c**, namely base point (at 0.25 V), peak point (at 0.50 V) and valley point (at 0.75 V). At the base point, well-localized scattering states are present on both electrodes and the additional overlap region (bottom panel of **Figure 4e**), due to lack of transport channels within the 0.25 V bias window around the $E_F$. At higher voltage (0.50 V), due to the additional electronic states in the channel evolving upon NO adsorption, there is a well-established connection throughout the device. Since the NO-induced electronic state is energetically localized, further increasing bias voltage starts to slowly break this connection and causes the current to drop (valley point).

From the transport and all other analyses presented in this work, we can again conclude that 1L or 2L SnP$_3$ could be a good sensing material, especially towards NO molecules. Moreover, combined with 3L or multi-layer SnP$_3$ as electrodes, results in a single-material logical junction with strongly reduced Schottky barrier.

**Summary.** We have investigated the potential of SnP$_3$ layered material for gas sensing applications. Its layered dependent electronic properties, namely 1L being semiconducting (used here as a transport channel) and 3L being metallic (used as electrodes), can be exploited for single-material devices, overcoming problems arising from the electrode-channels junction, such as high Schottky barrier. We investigated the adsorption of several common gases, $CH_4$, $CO_2$, $NH_3$, $NO_2$, and NO, as pollutants on $SnP_3$. Based on the adsorption energies, we categorized these molecules as type A ($CH_4$, $CO_2$, and $NH_3$) – weaker adsorption and type B ($NO_2$ and NO) – strong adsorption. The atomic charge analysis showed that type B has the largest charge transfer of about 0.7 $e^-$ from 1L $SnP_3$ to $NO_2$ and type A the smallest of 0.03 $e^-$ to $NH_3$. This is also visible in the projected density of states, where type B molecules hybridize with the state of $SnP_3$ close to the Fermi level. Next, we calculated transport properties for a device where 1L and 3L $SnP_3$ were used for transport channel and electrodes, respectively. We found significant changes in the transmission, T(E), due to the adsorption of type B molecules, consistent with the charge transfer and projected density of states analysis. While the current-voltage (*I-V*) characteristics for all molecules have similar shape up to 1.0 V bias, NO showed exceptional behavior, namely negative differential resistance (NDR) at about 0.6 V and 1.0 V. Thus, we analyzed the eigenchannels around the NDR of NO molecule and found that the scattering states, initially localized at each electrode, disperse throughout the device with increasing bias voltage and eventually establish a very good connection at the peak maximum (0.5 V). Further increasing the voltage starts to form "holes" in the scattering states leading to NDR. We also calculated recovery times of all molecules on 1L and 2L $SnP_3$ and found recovery time of 70 $\mu s$ for NO molecule on 2L $SnP_3$ substrate. Both NDR and very good recovery time reflect a great potential of $SnP_3$ for NO sensing application.

## ASSOCIATED CONTENT

**Supporting Information**. The Supporting Information contains additional results: electronic structures of monolayer, bilayer, and trilayer $SnP_3$; representation of different orientations of gas molecules adsorbed on the $SnP_3$ layer; the CDD plots for type A ($CO_2$ and $NH_3$) and type B ($NO_2$) molecules; transmission functions of pure $SnP_3$ in the 111- and 313-model; the *I-V* curve of pure $SnP_3$ in the 313-model.

## AUTHOR INFORMATION

**Corresponding Author**

Agnieszka B. Kuc – Department of Reactive Transport, Helmholtz-Zentrum Dresden-Rossendorf, 04318 Leipzig, German; orcid.org/0000-0002-9458-4136; Email: a.kuc@hzdr.de

Han Seul Kim – Center for Supercomputing Applications, National Institute of Supercomputing and Networking, Korea Institute of Science and Technology Information, 34141 Daejeon, Republic of Korea; Email: hanseulkim0@kisti.re.kr

**Author Contributions**

MSR performed the simulations. All authors contributed equally to analyzing the results and writing this manuscript.

**Funding Sources**

The authors declare no competing financial interest.
This research was supported by the Deutsche Forschungsgemeinschaft (project GRK 2247/1 (QM3)).

## ACKNOWLEDGMENT

Financial support by the Deutsche Forschungsgemeinschaft (GRK 2247/1 (QM3)). MSR and ABK acknowledge the high-performance computing center ZIH Dresden for computational resources. Association within SPP 2244 is acknowledged. HSK thank the Korea Institute of Science and Technology Information (KISTI) for high-performance computing resources (KSC-2020-CRE-0300) as well as the financial support (K-21-L02-C10), and the Ministry of Science and ICT of Korea for the research fund through National R&D program of the National Research Foundation (NRF) (NRF-2020R1F1A1075573).

# Gas Sensing Properties of single-material SnP3 logical junction via Negative Differential Resistance: Theoretical Study

## Supporting Information

**Muhammad S. Ramzan,**[†,‡] **Han Seul Kim,**[§, ¤,*] **Agnieszka B. Kuc,**[†,‡,*]

[†] Department of Physics and Earth Sciences, Jacobs University Bremen, Campus Ring 1, 28759 Bremen, Germany

[§] Center for Supercomputing Applications, National Institute of Supercomputing and Networking, Korea Institute of Science and Technology Information, 34141 Daejeon, Republic of Korea

[¤] Department of Data & High Performance Computing Science, University of Science & Technology, 34113 Daejeon, Republic of Korea

[‡] Helmholtz-Zentrum Dresden-Rossendorf, Abteilung Ressourcenökologie, Forschungsstelle Leipzig, Permoserstr. 15, 04318 Leipzig, Germany

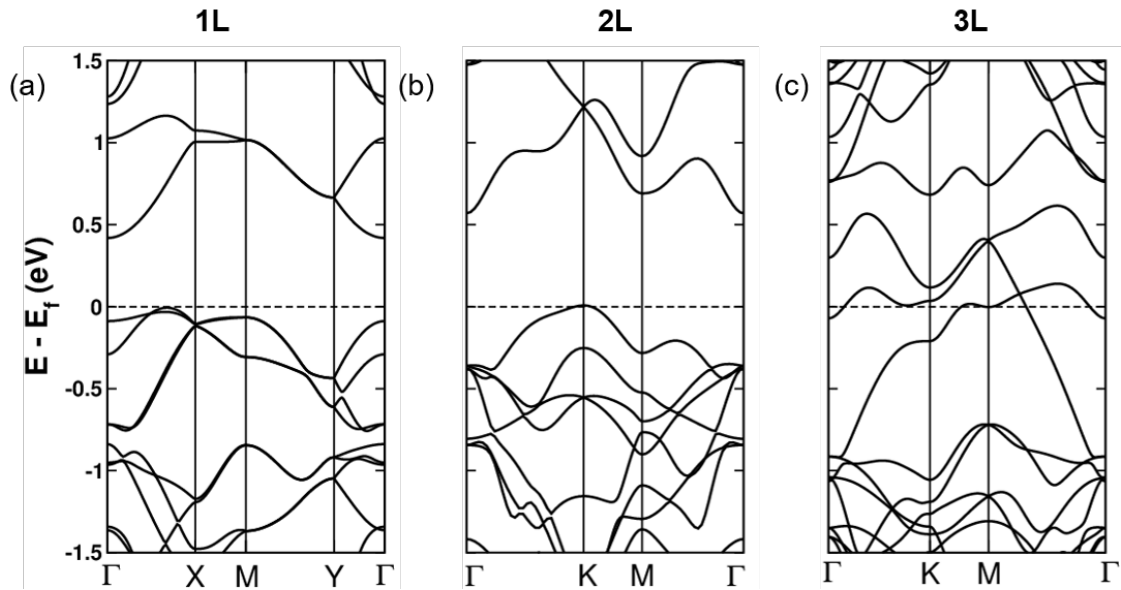

**Figure S1.** Electronic band structure of (a) monolayer, (b) bilayer and (c) trilayer, respectively, calculated at the PBE level of theory. Horizontal dashed lines reflect the Fermi levels which have been shifted to zero. The 1L band structure was calculated using a rectangular unit cell, while 2L and 3L using hexagonal unit cells. See Methods section in main text for details.

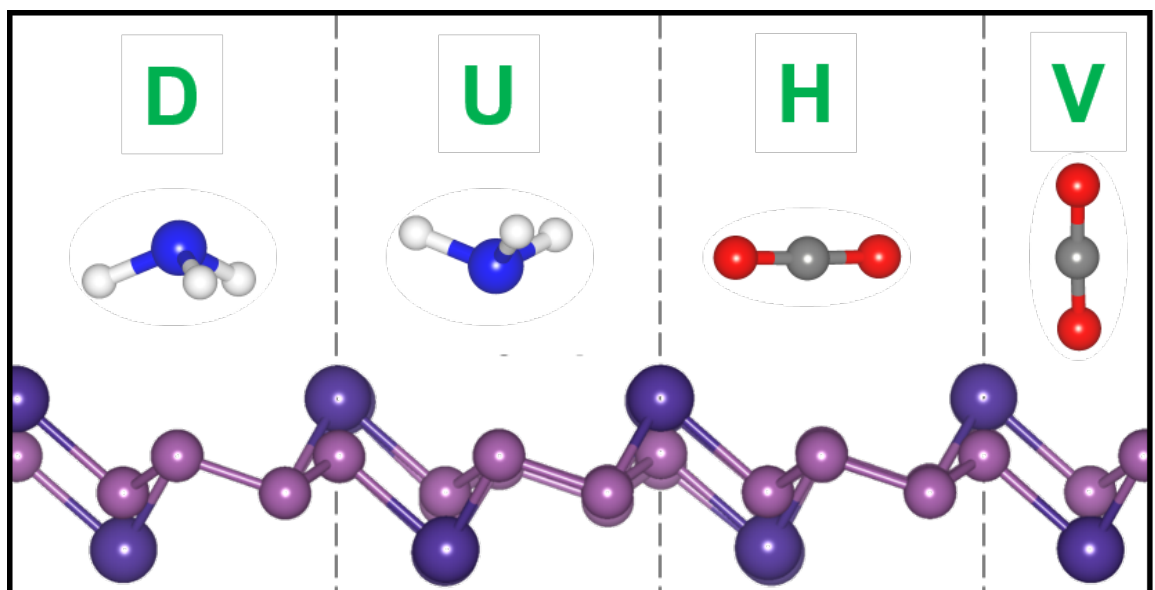

**Figure S2.** Side view of SnP$_3$ substrate with labelings for different adsorption molecular configurations, for planner and non-planner gas molecules, considered in this study.

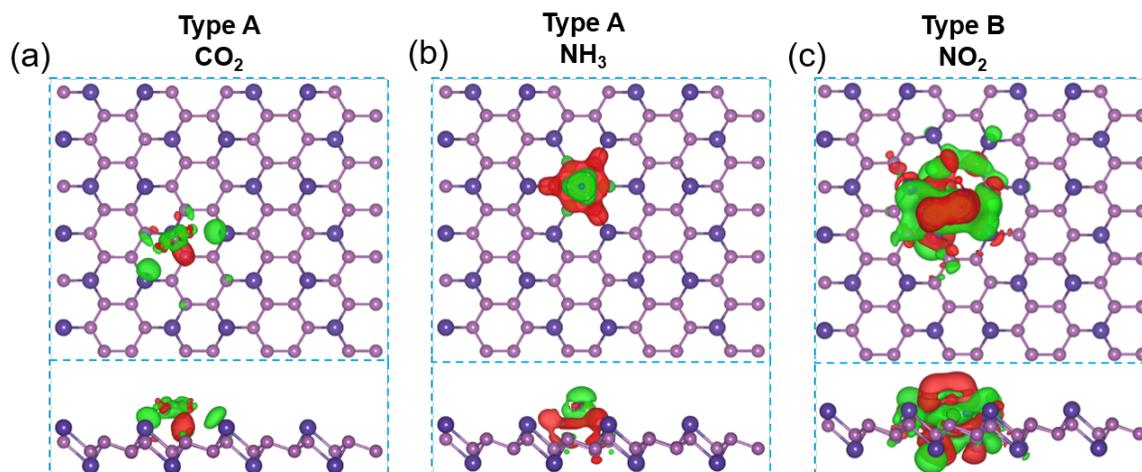

**Figure S3.** Charge density differences of type A (CO$_2$ and NH$_3$) and type B (NO$_2$) molecules. Green and red areas represent the charge accumulation and the charge depletion, respectively. The isovalue of $3 \times 10^4$ e/Å$^3$ is adopted.

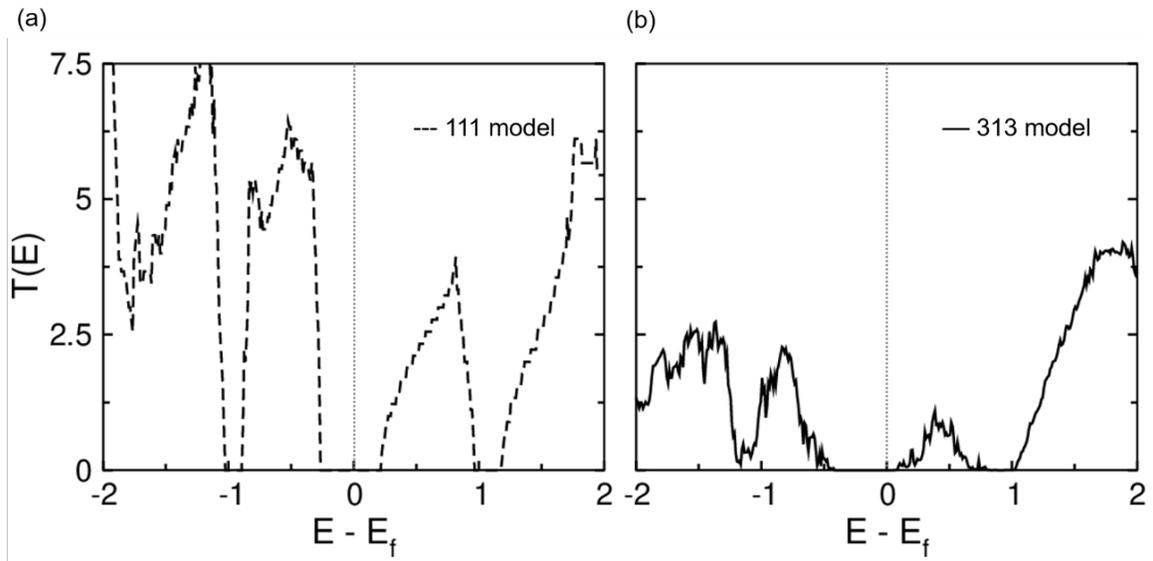

**Figure S4.** Transmission spectra of the pristine device using the 111- (a) and 313-model (b) with zero bias voltage. Vertical dashed lines at zero indicate Fermi levels.

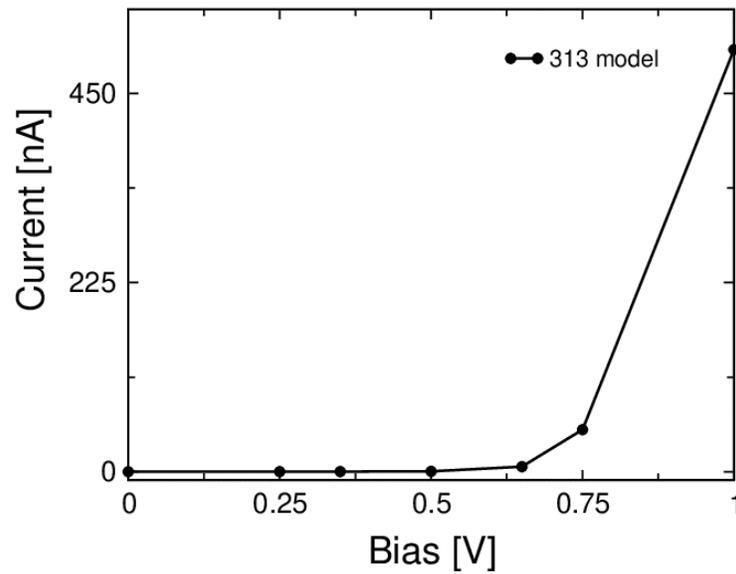

**Figure S5.** The current-voltage (*I-V*) characteristics of the pristine single-material device calculated using 313-model of SnP$_3$.

3